# Brillouin-derived viscoelastic parameters of hydrogel tissue models

Michelle Bailey,† Martina Alunni-Cardinali,# Noemi Correa,† Silvia Caponi,§ Timothy Holsgrove,^ Hugh Barr,¶ Nick Stone,† C. Peter Winlove,† Daniele Fioretto,#* and Francesca Palombo†*

† University of Exeter, School of Physics and Astronomy, Exeter EX4 4QL, UK

# University of Perugia, Department of Physics and Geology, Perugia I-06123, Italy

§ CNR-IOM - Istituto Officina dei Materiali - Research Unit in Perugia, c/o Dept. of Physics and Geology, University of Perugia, Perugia I-06123, Italy

^ University of Exeter, School of Engineering, Exeter EX4 4QF, UK

¶ Gloucestershire Royal Hospital, Gloucester, GL1 3NN, UK

* Corresponding authors: f.palombo@exeter.ac.uk, daniele.fioretto@unipg.it

**Short abstract**

Many problems in mechanobiology urgently require characterisation of the micromechanical properties of the fibrous proteins of cells and tissues. Brillouin light scattering has been proposed as a new optical elastography technique to meet this need, but the information contained in the Brillouin spectrum is still a matter of debate. Here we investigate this question using gelatin as a model system in which the macroscopic physical properties can be manipulated to mimic all the relevant biological states of matter, ranging from the liquid to the gel and the glassy phase. We demonstrate that Brillouin spectroscopy is a powerful tool to reveal both the elastic and viscous properties of biomaterials that are central to their biological functions.

**Long abstract**

The mechanical properties of cells and tissues are critical to their normal function, and alterations are linked to disease. The rapid advance of mechanobiology raises questions about mechanical properties on a microscopic scale. A novel optical elastography technique based on Brillouin spectroscopy promises such information, but there are fundamental problems in relating the data to low-frequency macroscopic mechanical parameters. In addition, biological tissues are complex hierarchical structures whose mechanical properties span a range of length and time scales, and this poses particular difficulties in establishing the structural and mechanical bases of the Brillouin signal. Here we use gelatin (denatured collagen) gels as model systems to evaluate the biophysical significance of the information provided by Brillouin spectroscopy. The data show that at low polymer concentration the spectrum is influenced by changes in the effective viscosity of solvent water arising from its interaction with the polymer and to the degree of cross-linking of the polymer matrix itself. Increasing the polymer concentration gives rise to spectral changes that reflect the longer time-scale dynamics: the sol–gel transition, the structural relaxation of the fluid phase and the eventual transition to the glassy phase that is accompanied by a fourfold increase in longitudinal Brillouin modulus.

In summary, our results show that if water plays a vital role in the spectroscopically determined viscoelastic properties, it is equally true that the Brillouin signal is also strongly influenced by the concentration and degree of cross-linking in the polymer phase. It is therefore clear that Brillouin spectroscopy is a powerful tool to probe many aspects of the physical environment in cells and tissues that are relevant to their biological functions and may be indicators of pathological dysfunctions.

**Introduction**

The macroscopic mechanical properties of the extracellular matrix are essential to the physiological functions of most biological tissues and on a microscopic scale they determine many aspects of cellular activity.[1-3] These properties are largely determined by networks of protein fibres, stiff fibres composed of 29 varieties of collagen and highly compliant ones composed primarily of elastin. Classical mechanical testing has provided a basis of understanding of how the composition and organisation of the networks in specific tissues yield the requisite mechanical properties and has demonstrated functionally significant changes in diseases ranging from osteoarthritis to atherosclerosis. However, research interest in these diseases has now moved to the sub-cellular level and this has generated an urgent need to characterise the mechanical properties of the extracellular matrix on these length scales. In this framework, Brillouin spectroscopy has emerged as a potential tool. Brillouin light scattering (BLS) is an acoustic process arising from the interaction of light with thermally driven acoustic phonons at high frequencies.[4,5]

Early work determined that this technique could measure the components of the elasticity tensor in dry collagen and elastin fibres[6-10] and showed that the longitudinal modulus, which is more easily measured in most scattering geometries, is many orders of magnitude higher than that determined by more classical engineering approaches. This discrepancy is presumed to reflect the different frequency scales of the two types of measurement, as well as that between the longitudinal modulus and the shear and Young's moduli that are more widely used in bioengineering. This distinction has been considered in recent literature.[11-13] However, a further complicating factor is the contribution of water both to fibre mechanics and to the Brillouin spectrum. The former is well established through the use of pore-elastic models to describe the mechanics of the extracellular matrix.[14,15] The latter is still a subject of controversy with some reports that in highly hydrated gels, simulating some aspects of the extracellular matrix, the frequency shift of the Brillouin peak is determined by modes generated in the water phase.[16]

Against this background we sought to explore the information content of the Brillouin spectrum generated by gelatin (denatured type-I collagen) gels, which are simple model systems displaying some of the pertinent mechanical properties of the extracellular matrix. By varying the polymer concentration it is possible to cover a range of static and dynamic macroscopic mechanical moduli that replicate those of many biological tissues. We show that for low polymer concentrations, the Brillouin linewidth, which is a viscosity indicator, is much more sensitive to concentration than frequency shift, thus making a full band shape analysis necessary to assess viscoelasticity in biological samples. We further show that the local increase in viscosity of the gelatin–water liquid phase is mainly due to a factor-1.9 retardation of the dynamics of water at the interface with gelatin, similar to the effect of hydrophobic hydration for a large class of biomimetic molecule mixtures. By decreasing the water content, we observe a liquid–glass transition that drives the system towards the solid-like behaviour typical of many tissues such as tendon and bone. Therefore covering a broad range of microenvironments that are significant for biological applications of Brillouin elastography.

These results demonstrate that with appropriate peak analysis Brillouin spectroscopy could become a probe of the micromechanics of biological tissues with important applications both in fundamental research and in clinical diagnosis.

**Results and Discussion**

*<u>1. High hydration.</u>* Brillouin spectra of collagen gels contain only a single, sharp peak (with Stokes and anti-Stokes components; see Fig. 1A), indicating that the gels are homogenous on the phonon wavelength scale (ca. 0.3 μm for 532 nm excitation). With increasing polymer concentration there is an increase in both the Brillouin frequency shift $\omega_B$ and linewidth $\Gamma_B$ (Fig. 1B), in line with previous works.[17,18] The values of $\omega_B$ and $\Gamma_B$ were derived from fitting the peaks to a damped harmonic oscillator (DHO) model [see Supplementary Information Fig. SI-1] (Fig. 1C). Since the ratio of density-to-refractive index square was approximately constant in all the investigated systems (Lorentz-Lorenz equation) [see Methods and Supplementary Information Fig. SI-2], the changes in $\omega_B$ and $\Gamma_B$ could be attributed unambiguously to an increase in the storage and loss moduli (Fig. 1D).

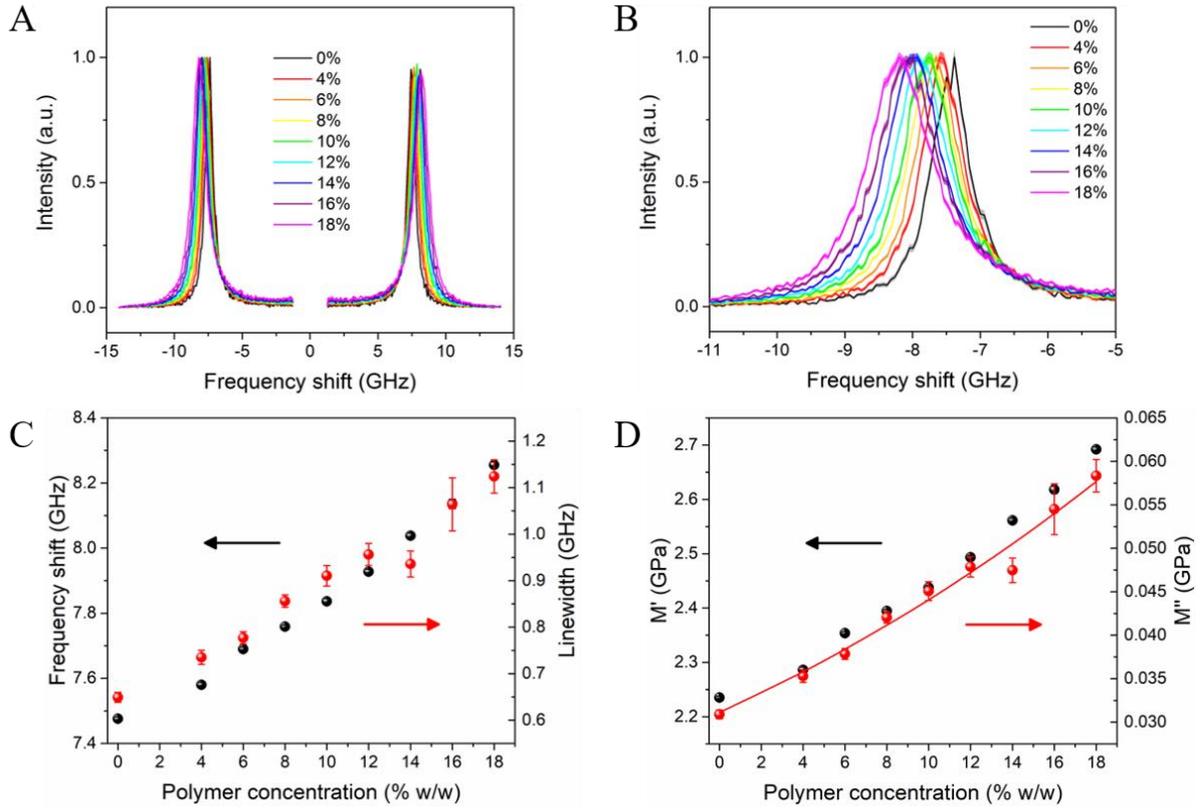

**Fig. 1** Dependencies of (A) Brillouin spectra, (B) Stokes peak, (C) frequency shift and linewidth, and (D) storage and loss moduli of gelatin gels on polymer concentration. Spectra are normalised to the maximum of the Stokes peak (B). Full circles: experimental data; error bars indicate the standard error (square root of number of counts). Red line: results of the linearized model (see below).

Changing polymer concentration has the largest effect on the width of the Brillouin peak (89% variation; Fig. 1C-D) which reflects variations in the microscopic viscosity of the system. This variation is attributed to the restricted mobility of water particularly in the first hydration shell of the collagen molecules (see Methods). This mechanism is shown in the schematic diagram of the frequency dependence of the storage ($M'(\omega)$) and loss ($M''(\omega)$) moduli at different polymer concentrations (Fig. 2). In the dilute polymer limit (Fig. 2a), the modulus at Brillouin frequencies (shadowed area) is that of a simple liquid, with a "relaxed" storage modulus $M_0' = \rho c_0^2$, where $c_0 = \frac{\omega_B}{q}$ is the adiabatic sound velocity, and a loss modulus $M'' = \omega_B b$, where $b = \rho \frac{\Gamma_B}{q^2}$ is the longitudinal kinematic viscosity ($\rho$: sample's mass density; $q$: exchanged wavevector, $q = 4\pi n/\lambda$ in backscattering geometry; $n$: sample's refractive index; $\lambda$: excitation wavelength, 532 nm).

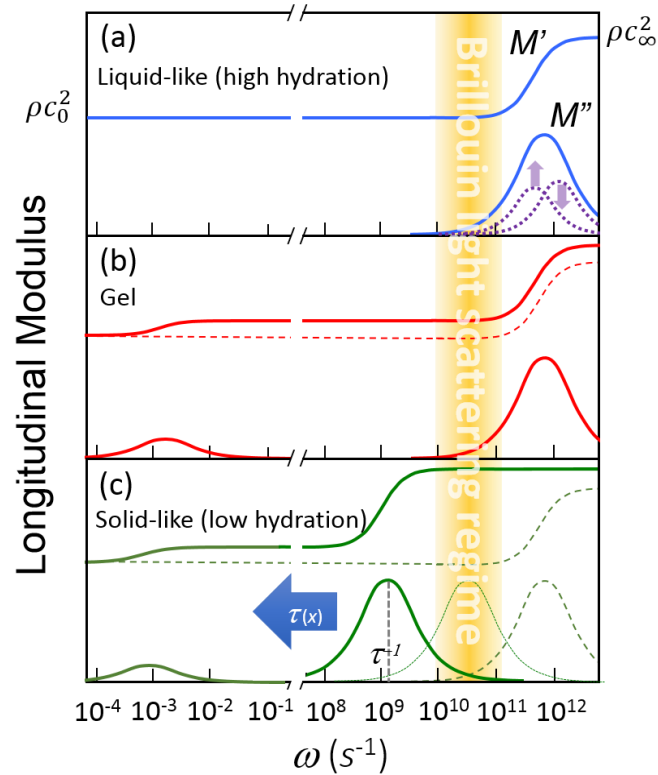

**Fig. 2** Schematic diagram of the dispersion in longitudinal storage modulus ($M'(\omega)$, upper curves) and loss modulus ($M''(\omega)$, lower curves) in the (a) high hydration limit, (b) gel phase and (c) low hydration limit. Yellow shaded area denotes the Brillouin region. (See also ref 5.)

Whilst the system remains in a liquid-like state, the structural relaxation responsible for the increase in modulus up to the solid-like value $M_\infty' = \rho c_\infty^2$ occurs on the picosecond time scale, i.e. at much higher frequency than the Brillouin peak, giving rise to a linear relationship between the loss modulus and frequency, whose gradient is given by the viscosity $b$, i.e. $M''(\omega) = \rho \omega b$. In the case of very diluted polymer, the water may be considered as two phases, free water, or "bulk" water, and interacting more strongly with the polymer chains, the "hydration" water. In this state, the observed increase of $\Gamma_B$ or $M''$ with increasing polymer concentration (Fig. 1C-D) can be attributed to an increase of *b* due to two contributions to the relaxation process (dotted lines in Fig. 2a), one from bulk and the other (retarded) from hydration water, which increases linearly with polymer concentration. Using this model and a hydration number derived from simple geometric arguments (10,071; see Methods), we obtain an average retardation of the order of 1.9 (Fig. SI-3) and values of $M''$ that match the experimental data (red line in Fig. 1D). This value of the retardation agrees with previously reported measurements and theoretical analyses of hydrophobic hydration that are appropriate to the structure of the collagen molecule.[19]

The evolution in storage modulus at constant temperature (Fig. 1D) shows only a 20% increase compared to 89% for the loss modulus. This is to be expected because the water modulus is high (2.2 GPa) and the addition of polymer produces only a small effect that, based on the rule of mixing, can be modelled as a weighted average of solvent and solute moduli (Fig. 3A). A fit to the Voigt model[20] across a broad range of concentrations (from pure water to 70% water content), yields a solute modulus of 5.64 GPa which is plausible for a highly hydrated network of collagen molecules. The Voigt model was found to fit the data better than the inverse relation (Reuss) used in previous works.

This result shows that in this system the Brillouin frequency shift is sensitive to the presence of the polymer network, and furthermore that the modulus of the network can be determined provided that the spectrometer provides adequate resolution.

Importantly, the Brillouin frequency shift is sufficiently sensitive to the elastic properties of the network to reveal the onset of a sol–gel transition which arises from the development of a percolative cluster involving a population of mutually linked collagen molecules. We observed this phenomenon by investigating two different gelatin concentrations, 10 and 20%, as a function of temperature. The sol–gel transition gives rise to a small "step" gradient of the frequency shift (see arrows in Fig. SI-4A). In contrast, the change in linewidth shows no discontinuity at the gel transition point (Fig. SI-4B). This is consistent with the fact that even in the gel phase a major fraction of molecules are still in the liquid phase and their diffusive motion dominates the picosecond dynamics giving rise to the broadening of the Brillouin spectra in both the liquid and gel phase. These motions are arrested at the glass transition. The effect of a sol–gel transition on the loss and storage moduli is schematically depicted in Fig. 2b for a polymer concentration high enough (or temperature low enough) to make a transition to the gel state. The solid-like portion of the sample, composed of cross-linked collagen molecules, is responsible for the relaxation process occurring at very long time scales (hundreds of seconds or more), and this gives rise to the divergence of the "static" viscosity and the onset of shear ($G$) and Young's ($E$) moduli. The small number of molecules involved in this process is responsible for two phenomena:

1) A state in which the values of $G$ and $E$ are orders of magnitude smaller than the longitudinal modulus. Note that the Young's moduli derived from compressive testing on these hydrogels are of the order of kPa (Fig. SI-5), whilst the high-frequency longitudinal moduli are in the GPa range;

2) A small jump in the value of $M'$ (from dashed to solid line in Fig. 2b) revealed by the experiments reported in Fig. SI-4A and smooth transition in $M''$ as shown in Fig. SI-4B because the liquid fraction of the sample is almost unchanged by the gelation process and is still responsible for the high frequency (picosecond) relaxation, which is comparable to that of the diluted solutions (Fig. 2a).

*2. Low hydration.* As the water fraction in the samples is reduced, the dynamics of the remaining water are increasingly closely coupled to those of the collagen molecules until an arrested (glassy) phase is attained. It is worth noting that this is the mechanism of hardening of "animal glue", one of the most widely used glues worldwide. Indeed, the name "collagen" derives from the Greek "kolla", glue. The transition from the liquid phase (low elastic modulus) to the solid phase (high elastic modulus) is revealed in the frequency dispersion and the associated maximum in linewidth in the Brillouin data (Fig. 3B).

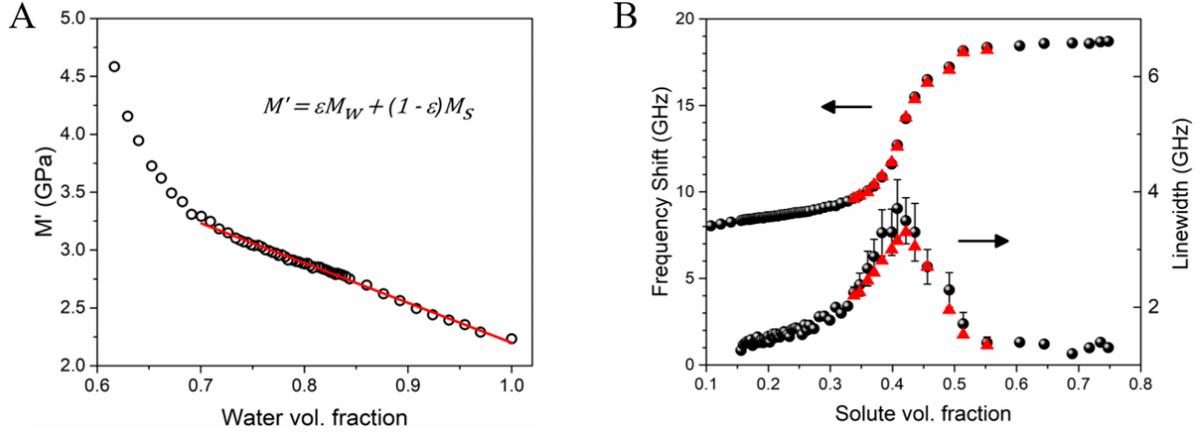

**Fig. 3** Evolution in (A) storage modulus vs. water volume fraction, $\varepsilon$. Linear fit to a Voigt model (Eq. in figure) yields the elastic moduli of water and solute: $M_w$= 2.20 GPa and $M_s$= 5.64 GPa; $R^2$=0.994. (B) (Upper curve) Brillouin frequency shift and (lower curve) linewidth of gelatin vs. polymer concentration. The error bars encompass the range of values obtained by the fits. The red triangles denote the frequency shifts and linewidths of the theoretical curves obtained by the viscoelastic fit.

The glass transition in dried collagen was first investigated through a more traditional thermodynamic path by Flory and Garret[21] and attributed to the temperature-induced arrest of the amorphous fraction and of the side chains of collagen molecules. As in analyses of the hardening of epoxy resins,[22,23] the Mode Coupling Theory (MCT)[24] provides a rationale for the early stage of the structural arrest associated with the glass transition of collagen (see Methods).

Fig. 2c qualitatively shows the effect of progressive dehydration as the structural relaxation shifts towards lower frequencies. As deduced by Eq3 (see Methods), the condition of maximum linewidth is reached when the maximum of $M''(\omega)$ matches the frequency of the Brillouin peak (dotted curve in Fig. 2c). Further reduction of hydration (full line in Fig. 2c) gives rise to a reduction of linewidth and an increase of Brillouin frequency shift approaching the solid-like (unrelaxed) condition $M_\infty' = \rho c_\infty^2$. The full viscoelastic treatment of Brillouin spectra in this regime, the early stage of liquid–glass transition, by means of the Mode Coupling Theory is reported in Methods, giving the red triangles in Fig. 3B. The signatures of singularity in the non-ergodicity parameter $(1 - c_0^2/c_\infty^2)$ and of divergence of the structural relaxation (see Fig. 4 below), consistent with the predictions of MCT, suggest that the ergodicity breakdown described by MCT is a more "universal" mechanism of glass transition than ever supposed to be before.

In summary, Brillouin scattering has revealed:

- High sensitivity to small changes in elastic modulus in diluted samples (gelation);
- Large changes in loss modulus in diluted samples due to hydrophobic hydration;
- Relatively small increase of micro-viscosity up to 30-40% polymer concentration when compared to static viscosity, which already diverged due to gelation, suggesting the possibility of diffusion of small molecules in already macroscopically solid-like samples;
- Early stage of microscopic transition from liquid to solid state at ca. 47% polymer concentration, well described by the Mode Coupling Theory of glass transition.

**Methods**

*<u>Hydrogel preparation.</u>* Hydrogels were prepared as previously described,[25] from ~225 g Bloom type-B bovine skin gelatin powder (G9382, Sigma-Aldrich) and distilled water in the concentration range 4–18 % w/w gelatin. Gelatin powder and water were mixed under magnetic stirring, while being heated in a water bath at 55–65 °C for 60 minutes, to ensure complete dissolution. Gels were left to cool to 40°C and were then transferred to storage vessels (see below), sealed to reduce evaporation effects and then left to cool down to room temperature. Measurements were conducted approximately 24 hours after preparation, as preliminary testing established that this time was sufficient for the gel to stabilise.

*<u>Refractive index and density measurements.</u>* The refractive index of all samples prepared was measured using an Abbe refractometer (Atago model NAR-1T; resolution 0.0002 nD), with distilled water as a reference. A small sample of gelatin was removed from the bulk prior to gelation (at ~40°C) and left to set in between the plates of the refractometer, ensuring good contact with both plates. Measurements were taken when the gel reached room temperature (Fig. SI-2A).

Density was determined from a biphasic ideal mixing model:

$$\rho = \frac{(m_w + m_g)}{\left(\frac{m_w}{\rho_w} + \frac{m_g}{\rho_g}\right)} \tag{1}$$

where $m_w$ and $m_g$ denote the mass of water and dry gelatin respectively, and $\rho_w$ and $\rho_g$ the mass densities. Densities were taken to be 1.00 g/cm$^{-3}$ for water, and 1.35 g/cm$^3$ for dry gelatin.[26]

It is observed that the change in density-to-square refractive index ratio is less than 1% across the concentration range probed here (Fig. SI-2B).

*Compressive testing.* After preparation, the gels were transferred into custom-built aluminium moulds. An Instron ElectroPuls E10000 (High Widcombe, UK) linear dynamic test instrument was used to perform unconfined compressive testing on the hydrogels at a steady rate of 0.03 mm/s. The cylindrical gel sample (21 mm diameter, ~10 mm high) sat on a flat aluminium base and the applied load was measured by a 1 kN load cell (<2% linearity error) mounted onto a flat aluminium plate in contact with the top surface of the gel. Applied strain of up to 10% did not cause a significant change in shape of the gels radially. Hence from the uniaxial deformation, the gradient of the stress–strain curve was used to calculate the Young's modulus of the hydrogel samples (Fig. SI-5).

*Brillouin spectroscopy.* Brillouin spectra of the hydrogels were acquired using a high-contrast tandem Fabry-Pérot (TFP-2 HC) interferometer system, with 532 nm cw laser and a 20× (NA 0.42) Mitutoyo objective, as previously described.[27] Spectral resolution was 135 MHz and contrast >150 dB. Laser power at the sample was a few mW and acquisition time 17 s per spectrum.

*High hydration.* Hydrogels prepared up to the solubility limit (18%) were transferred to glass vials and measurements were conducted in triplicate using a backscattering geometry with 180° scattering angle. At two concentrations, 10 and 20% gelatin, measurements were also conducted as a function of temperature, from 65 to 4-5°C (water bath), to encompass the sol–gel transition.

*Low hydration.* For hydrogel dehydration, a thin film (200-300 μm thickness) was deposited onto a reflective silicon substrate and positioned on an analytical scale (Sartorius BL 210 S) to monitor the change in concentration. Measurements were conducted according to a previously established protocol,[9,10] using an achromatic lens (NA 0.033), and the sample was positioned at a 45° angle to the incident beam with a laser power of ~35 mW. Both bulk and parallel-to-surface acoustic modes were detected in this geometry [see Supplementary Methods in SI and Fig. SI-6].

Brillouin peaks at high hydration were analysed using a damped harmonic oscillator (DHO) function[5,28] in the range 7–9 GHz. Average fit parameters from Stokes and anti-Stokes peaks were obtained after deconvolution of the instrumental response function. Note that these gels present negligible multiple scattering effects;[29] indeed repeating the fit analysis in the range 8–13 GHz, taking only a small portion of the low-frequency side of the peaks, gave <0.1% difference in frequency shift and <1% difference in linewidth compared to the other fit.

*Longitudinal elastic moduli.* The storage modulus $M'$ was derived from the Brillouin frequency shift $\omega_B$ through the relation:

$$M' = \left(\frac{\lambda}{4\pi}\right)^2 \frac{\rho}{n^2} \omega_B^2 \qquad (2)$$

where $\frac{\lambda \omega_B}{4\pi n} = c$ is the acoustic wave velocity, $\lambda$ the excitation wavelength (532 nm), $\rho$ and $n$ the mass density and refractive index of the medium. Eq2 shows that there is a direct relation between the real part of the longitudinal modulus and the square Brillouin shift, through the ratio $\rho/n^2$.

In a similar way, the loss modulus $M''$ can be derived from the Brillouin frequency shift $\omega_B$ and linewidth $\Gamma_B$:

$$M'' = \left(\frac{\lambda}{4\pi}\right)^2 \frac{\rho}{n^2} \omega_B \Gamma_B \qquad (3)$$

Note: Eqs2-3 are valid for backscattering geometry, which is typical of microscopy applications.

*Linearized model at high hydration.* In the case of water and diluted aqueous solutions, the frequency of Brillouin lines is much smaller than the molecular relaxation rates (Fig. 2a). In this "relaxed" condition, the linewidth of Brillouin peaks $\Gamma_B$ yields the "longitudinal" kinematic viscosity $b$ of the liquid through the relationship $b = \rho \Gamma_B/q^2$, where $q$ is the exchanged wavevector, $q = 4\pi n/\lambda$ in backscattering geometry.

The longitudinal kinematic viscosity, in turn, can give information on the characteristic times of molecular relaxations. For aqueous solutions, density fluctuations are generally characterized by a two-step relaxation associated to hydration and bulk water with characteristic times, $\tau_h$ and $\tau_b$ respectively. The longitudinal viscosity can then be written as:[19]

$$b = \Delta_c[\alpha \tau_h + (1-\alpha)\tau_b] + b_\infty \qquad (4)$$

where $\alpha$ is the fraction of relaxation strength associated to hydration water, $b_\infty$ accounts for contributions to viscosity that are very fast (instantaneous) relative to the picosecond time scale investigated by Brillouin scattering, and $\Delta_c$ is the total relaxation strength of the solution at a given polymer concentration, given by relaxed ($c_0$) and unrelaxed ($c_\infty$) sound velocities through the relationship: $\Delta_c = c_\infty^2 - c_0^2$. Assuming that the relaxation strengths of the two processes are proportional to the relative fractions of hydration and bulk water, we can express $\alpha = f_r N_h$, where $f_r$ is the fraction of polymer-to-water molecules and $N_h$ is the hydration number. Moreover, hydration water molecules are typically found to diffuse more slowly than bulk molecules, and the retardation parameter $\varepsilon = \tau_h/\tau_b$ is used to quantify this effect. Eq4 can thus be rearranged as:

$$\frac{\Delta_0}{\Delta_c}\frac{b-b_\infty}{b_0-b_\infty} - 1 = N_h(\varepsilon - 1)f_r \tag{5}$$

where $b_0$ and $\Delta_0$ are the kinematic viscosity and relaxation strength of pure water, with $c_0$ obtained from the frequency position of Brillouin peaks and $c_\infty$ = 2860 m/s measured by Inelastic X-ray Scattering (IXS)[30] and assumed to be independent of polymer concentration;[31] $b_\infty$ = 2.99 10$^{-3}$ cm²/s derived from IXS measurements.[30]

The prediction of Eq5 for the change of viscosity of the solution as a function of polymer concentration is tested in Fig. SI-3. A well-defined linear behaviour of $\frac{\Delta_0}{\Delta_c}\frac{b-b_\infty}{b_0-b_\infty} - 1$ vs. $f_r$ can be seen across the whole range of polymer molecular fraction and the linear fit of the data gives $N_h(\varepsilon - 1) = 9566$ ($\pm$ 290). We then used a combination of this result derived using a linearized model from hydrodynamic theory[19] and published data from MD simulations[32] to derive the retardation factor for this system. The hydration number $N_h$ can be estimated by geometrical arguments as the number of water molecules (number density $\rho$ = 33.37 10$^{27}$ m$^{-3}$) within 3.1 Å of the triple helix surface, by modelling the collagen triple helix as a rod of 0.36 nm radius and 300 nm length,[32] giving $N_h$ = 10k. From this value, a retardation factor $\varepsilon$ = 1.9 can be deduced for hydration water, which is in the range previously found for hydrophobic hydration in a large class of biomimetic molecules.[19,31] The close agreement between the prediction of our two-step relaxation model and the measured concentration dependence of Brillouin parameters, together with the adequate value obtained for the retardation of hydration water, supports the view that water forming the first hydration shell of collagen molecules has a great impact on the dynamics of these gelatins, thus affecting the viscosity and loss modulus far more than the elasticity and storage modulus.

*Glass transition.* The glass transition is a dynamic process that occurs as an abrupt increase of the structural relaxation time, leading the system out of equilibrium (ergodic to non-ergodic transition).[24] In the frame of MCT, the transition is induced by a slowdown of density fluctuations which, in the frequency domain, can be described by the complex frequency-dependent longitudinal modulus $M(\omega) = M'(\omega) + iM''(\omega)$. Close to the glass transition, a stretched exponential relaxation of the longitudinal modulus typically occurs, described by the Kohlrausch-Williams-Watts (KWW) law: $e^{(-t/\tau)^\beta}$, where $\tau$ is the characteristic time and $\beta < 1$ is the stretching parameter. In the frequency domain, the Fourier transform of the KWW law can be conveniently described by an Havriliak-Negami (HN) relaxation function:[33]

$$\frac{M(\omega)}{\rho} = c_\infty^2 - \frac{c_\infty^2 - c_0^2}{[1+(i\omega\tau)^a]^b} \tag{6}$$

where $c_0$ and $c_\infty$ are the relaxed (low frequency $\omega$, or low $\tau$ with respect to the Brillouin frequency, such as in the high hydration regime) and unrelaxed (high $\omega$, or high $\tau$, low hydration) sound velocities; $a$ and $b$ are the stretching parameters determined by the value of the KWW $\beta$ parameter.[33] The Mode Coupling Theory, in its basic formulation, predicts a power law divergence of the relaxation time $\tau \propto (-\varepsilon)^{-\gamma}$ and a square root singularity of the amplitude of the structural relaxation (non-ergodicity parameter) $1 - c_0^2/c_\infty^2 = f_q = f_q^c + h_q\sqrt{\varepsilon}$, for the control parameter $\varepsilon \to 0$. Depending on the experimental path, the control parameter can be expressed in terms of temperature (thermal vitrification) or density (pressure vitrification) or volume fraction of polymer molecules in case of colloidal suspensions.[24] Here we define it in terms of volume fraction of collagen molecules $x$ as: $\varepsilon = (x - x_0)/x_0$, where $x_0$ is the "ideal" critical concentration for the structural arrest. According to MCT, the values of the parameters regulating the stretching of the relaxation function and the power law divergence of the relaxation time depend only on the structure of the sample and are mutually related by well-defined analytical expressions.[34] Measuring these parameters give a quantitative test of the predictions of the theory.

Brillouin spectra from longitudinal acoustic modes are informative of $M(\omega)$ and can be used to test the predictions of the MCT, since they give direct access to the spectrum of density fluctuations (fluctuation-dissipation theorem)[35]:

$$I_q(\omega) = \frac{I_0}{\omega} \Im\{[M(\omega)/\rho - \omega^2/q^2]^{-1}\}. \tag{7}$$

This equation shows that the maximum of information (maximum intensity) in the Brillouin spectrum is at the resonance (Brillouin peak) occurring around $\omega_B = (q^2 M'(\omega_B)/\rho)^{1/2}$. Unfortunately, the fit of a single Brillouin spectrum to this equation is not sufficient to get the whole set of relaxation parameters $c_0$, $c_\infty$, $\tau$ and $\beta$. Different strategies can be implemented to mitigate this problem.[28] In the present work, we expanded the frequency range by collecting light from two simultaneous scattering geometries [see Supplementary Methods in SI] and we independently estimated the values of $n$, $c_0$ and $\beta$, so that $c_\infty$ and $\tau$ were the only free parameters in fitting Brillouin spectra.

From the fit, the concentration dependence of the relaxation time $\tau$ and of the non-ergodicity parameter $1 - c_0^2/c_\infty^2$ was obtained and reported in Fig. 4 to be compared with the predictions of the MCT.

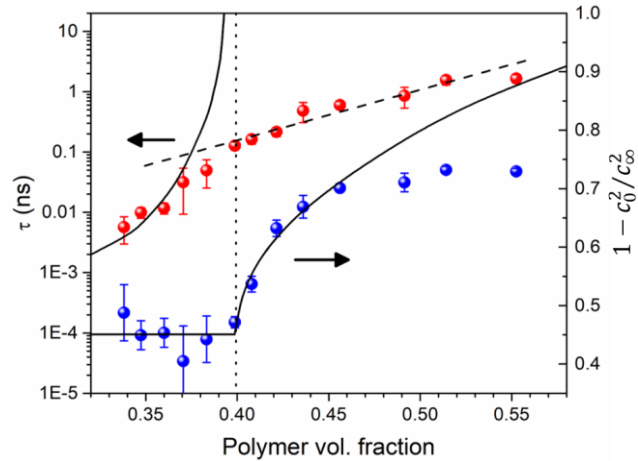

**Fig. 4** Plot of the relaxation time $\tau$ and non-ergodicity parameter $f_q = 1 - c_0^2/c_\infty^2$ vs. polymer volume fraction $x$. Deviations of data points (full circles) from ideal behaviour (solid lines) can be explained by secondary relaxation processes, which are highly likely in real systems. Dashed line is a guide for the eye; dotted line denotes the "ideal" critical concentration for the structural arrest predicted by the Mode Coupling Theory.

Though the system investigated here is by far more complex than the liquids and colloidal suspensions usually analysed with MCT, signatures of a critical concentration $x_0$ located around 40% polymer can be clearly seen in Fig. 4. In particular, $f_q$ shows an inflection close to $x_0$, with an increase at higher concentrations that mimics the square root behaviour predicted by the theory (solid line). Moreover, at lower concentrations, $\tau$ follows the power law behaviour predicted by the theory, with an exponent $\gamma$ of approximately 4, the value related by MCT to the stretching parameter of the structural relaxation $\beta = 0.45$ [see Supplementary Methods in SI]. The deviation from the power law visible in Fig. 4 when approaching the critical point is quite typical for all glass-forming systems[24] and attributed to the presence of additional (secondary) relaxation processes responsible for restoring ergodicity above $x_c$.

As a whole, these results are in good agreement with the predictions of the Mode Coupling Theory for the glass transition, which was previously verified in simple glass formers through more traditional thermodynamic paths, namely with temperature as the control parameter.[24] This was previously verified as a function of temperature[36,37] and pressure,[38] upon hardening of epoxy glues[23] and astonishingly here, for the first time (to our knowledge), in hardening of "kolla" controlled by water concentration.

# References

1  Vining, K. H. & Mooney, D. J. Mechanical forces direct stem cell behaviour in development and regeneration. *Nature Reviews Molecular Cell Biology* **18**, 728-742 (2017).
2  Swaminathan, V. *et al.* Mechanical stiffness grades metastatic potential in patient tumor cells and in cancer cell lines. *Cancer research*, canres.0247.2011.Can-11-0247 (2011).
3  Northcott, J. M., Dean, I. S., Mouw, J. K. & Weaver, V. M. Feeling Stress: The Mechanics of Cancer Progression and Aggression. *Frontiers in Cell and Developmental Biology* **6** (2018).
4  Brillouin, L. Diffusion de la lumière et des rayonnes X par un corps transparent homogène; influence de l'agitation thermique. *Ann. Phys.* **17**, 88-122 (1922).
5  Palombo, F. & Fioretto, D. Brillouin Light Scattering: Applications in Biomedical Sciences. *Chemical Reviews* **119**, 7833-7847 (2019).
6  Randall, J., Vaughan, J. M. & Cusak, S. Brillouin Scattering in Systems of Biological Significance [and Discussion]. *Philosophical Transactions of the Royal Society of London. Series A, Mathematical and Physical Sciences* **293**, 341-348 (1979).
7  Cusack, S. & Miller, A. Determination of the elastic constants of collagen by Brillouin light scattering. *Journal of Molecular Biology* **135**, 39-51 (1979).
8  Cusack, S. & Lees, S. Variation of longitudinal acoustic velocity at gigahertz frequencies with water content in rat-tail tendon fibers. *Biopolymers* **23**, 337-351 (1984).
9  Palombo, F. *et al.* Biomechanics of fibrous proteins of the extracellular matrix studied by Brillouin scattering. *J. R. Soc. Interface* **11**, 12 (2014).
10  Edginton, R. S. *et al.* Preparation of Extracellular Matrix Protein Fibers for Brillouin Spectroscopy. *Journal of Visualized Experiments*, e54648 (2016).
11  Edginton, R. S., Green, E. M., Winlove, C. P., Fioretto, D. & Palombo, F. in *SPIE BiOS.* 7 (SPIE).
12  Caponi, S., Canale, C., Cavalleri, O. & Vassalli, M. in *Nanotechnology Characterization Tools for Tissue Engineering and Medical Therapy* (ed C. S. S. R. Kumar) Ch. 2, 69-111 (Springer-Verlag GmbH 2019).
13  Prevedel, R., Diz-Muñoz, A., Ruocco, G. & Antonacci, G. Brillouin microscopy: an emerging tool for mechanobiology. *Nature Methods* **16**, 969-977 (2019).
14  Mow, V. C., Wang, C. C. & Hung, C. T. The extracellular matrix, interstitial fluid and ions as a mechanical signal transducer in articular cartilage. *Osteoarthritis and Cartilage* **7**, 41-58 (1999).
15  Frantz, C., Stewart, K. M. & Weaver, V. M. The extracellular matrix at a glance. *J Cell Sci* **123**, 4195-4200 (2010).
16  Adichtchev, S. V. *et al.* Brillouin spectroscopy of biorelevant fluids in relation to viscosity and solute concentration. *Phys. Rev. E* **99**, 062410 (2019).
17  Bot, A., Schram, R. P. C. & Wegdam, G. H. Brillouin light scattering from a biopolymer gel: hypersonic sound waves in gelatin. *Colloid and Polymer Science* **273**, 252-256 (1995).
18  Bedborough, D. S. & Jackson, D. A. Brillouin scattering study of gelatin gel using a double passed Fabry-Perot spectrometer. *Polymer* **17**, 573-576 (1976).
19  Comez, L., Lupi, L., Paolantoni, M., Picchio, F. & Fioretto, D. Hydration properties of small hydrophobic molecules by Brillouin light scattering. *J Chem Phys* **137**, 114509 (2012).

20  Clyne, T. W. & Withers, P. J. in *An Introduction to Metal Matrix Composites Cambridge Solid State Science Series* 12-43 (Cambridge University Press, 1993).
21  Flory, P. J. & Garrett, R. R. Phase Transitions in Collagen and Gelatin Systems1. *Journal of the American Chemical Society* **80**, 4836-4845 (1958).
22  Corezzi, S., Fioretto, D. & Rolla, P. Bond-controlled configurational entropy reduction in chemical vitrification. *Nature* **420**, 653 (2002).
23  Corezzi, S., Comez, L., Monaco, G., Verbeni, R. & Fioretto, D. Bond-Induced Ergodicity Breakdown in Reactive Mixtures. *Phys. Rev. Lett.* **96**, 255702 (2006).
24  Gotze, W. & Sjogren, L. Relaxation processes in supercooled liquids. *Reports on Progress in Physics* **55**, 241-376 (1992).
25  Kalyanam, S., Yapp, R. D. & Insana, M. F. Poro-viscoelastic behavior of gelatin hydrogels under compression-implications for bioelasticity imaging. *Journal of biomechanical engineering* **131**, 081005 (2009).
26  Taffel, A. CCXXXVI.—Thermal expansion of gelatin gels. *Journal of the Chemical Society, Transactions* **121**, 1971-1984 (1922).
27  Scarponi, F. *et al.* High-Performance Versatile Setup for Simultaneous Brillouin-Raman Microspectroscopy. *Physical Review X* **7**, 031015 (2017).
28  Comez, L., Masciovecchio, C., Monaco, G. & Fioretto, D. in *Solid State Physics* Vol. 63 (eds E. Camley Robert & L. Stamps Robert) 1-77 (Academic Press, 2012).
29  Corezzi, S., Comez, L. & Zanatta, M. A simple analysis of Brillouin spectra from opaque liquids and its application to aqueous suspensions of poly-N-isopropylacrylamide microgel particles. *Journal of Molecular Liquids* **266**, 460-466 (2018).
30  Monaco, G., Cunsolo, A., Ruocco, G. & Sette, F. Viscoelastic behavior of water in the terahertz-frequency range: An inelastic x-ray scattering study. *Phys. Rev. E* **60**, 5505-5521 (1999).
31  Lupi, L. *et al.* Hydrophobic hydration of tert-butyl alcohol studied by Brillouin light and inelastic ultraviolet scattering. *J Chem Phys* **134**, 055104 (2011).
32  Varma, S., Orgel, J. P. & Schieber, J. D. Nanomechanics of Type I Collagen. *Biophys J* **111**, 50-56 (2016).
33  Alvarez, F., Alegra, A. & Colmenero, J. Relationship between the time-domain Kohlrausch-Williams-Watts and frequency-domain Havriliak-Negami relaxation functions. *Physical Review B* **44**, 7306-7312 (1991).
34  Gotze, W. The scaling functions for the β-relaxation process of supercooled liquids and glasses. *Journal of Physics: Condensed Matter* **2**, 8485-8498 (1990).
35  Montrose, C. J., Solovyev, V. A. & Litovitz, T. A. Brillouin Scattering and Relaxation in Liquids. *The Journal of the Acoustical Society of America* **43**, 117-130 (1968).
36  Caponi, S. *et al.* Ergodicity breaking in strong and network-forming glassy systems. *Physical Review B* **79**, 172201 (2009).
37  Mallamace, F. *et al.* On the ergodicity of supercooled molecular glass-forming liquids at the dynamical arrest: the o-terphenyl case. *Scientific Reports* **4**, 3747 (2014).
38  Casalini, R., Paluch, M. & Roland, C. M. Dynamic crossover in supercooled liquids induced by high pressure. *Journal of Chemical Physics* **118**, 5701-5703 (2003).

**Acknowledgements**

The authors would like to thank Dr Ellen M. Green, Emily Francis-Pollin and Morgan Nancarrow at the University of Exeter for stress–strain testing and helpful discussions. This work was supported by the UK Engineering and Physical Sciences Research Council (EP/M028739/1) and jointly by Cancer Research UK (NS/A000063/1). MB was also supported by the EU COST Action BioBrillouin (CA16124) for a Short Term Scientific Mission at DF's lab.

**Author contributions**

FP, DF and CPW conceived, designed and supervised the project. SC and DF supervised the work in the Brillouin Lab. SC, MB and MAC performed all experiments. MB, FP and DF processed and analysed the data. TH and CPW helped with compressive testing analysis and discussion of the results. MB, FP and DF wrote the manuscript with input from all other authors.